# Optoelectronic devices, plasmonics and photonics with topological insulators

Antonio Politano[1,] Leonardo Viti[2], and Miriam S. Vitiello[2]

[1]Department of Physics, University of Calabria, via ponte Bucci, cubo 31/C 87036, Rende (CS), Italy

[2]NEST-Istituto Nanoscienze and Scuola Normale Superiore, Piazza San Silvestro 12, 56127 Pisa, Italy

**Abstract**

Topological insulators are innovative materials with semiconducting bulk together with surface states forming a Dirac cone, which ensure metallic conduction in the surface plane. Therefore, topological insulators represent an ideal platform for optoelectronics and photonics. The recent progress of science and technology based on topological insulators enables the exploitation of their huge application capabilities. Here, we review the recent achievements of optoelectronics, photonics and plasmonics with topological insulators. Plasmonic devices and photodetectors based on topological insulators in a wide energy range, from Terahertz to the ultraviolet, promise outstanding impact. Furthermore, the peculiarities, the range of applications and the challenges of the emerging fields of topological photonics and thermoplasmonics are discussed.

**Introduction**

Topological insulators (TIs) represent a novel quantum phase of matter [1], characterized by a semiconducting bulk and topologically protected surface states [2-5].

Topological surface states (TSS) form a Dirac cone[6, 7], in analogy with graphene . In addition, TSS are chiral [8] and protected from back-scattering by time-reversal symmetry [9]. Therefore, TSS are robust against scattering processes from nonmagnetic impurities [10]. As a consequence of the time-reversal symmetry, charge carriers from topologically protected surface states carry current with minimal dissipation with subsequent reduction of the low-frequency electronic noise [11, 12].

Moreover, unlike graphene, TSS naturally offer a two-dimensional (2D) Dirac fermion system [13], without the complication of the physical implementation of atomically thin layers.

TIs can be divided into two classes: three-dimensional (3D) [14] and 2D [15] TIs. TSS are present in the bulk energy gap of 3D TIs, while 2D TIs consist of 1D gapless conductive edge channels with a 2D area exhibiting an energy gap.

Many classes of materials have TI properties: the mercury- cadmium telluride HgTe/CdTe quantum wells [16] are 2D TIs, whereas the bismuth[5, 17] or antimony[18, 19] chalcogenides, ternary chalcopyrites[20], the monoclinic phase of silver telluride ($\beta$-$Ag_2Te$)[21] etc. are 3D TIs.

The crystalline quality of samples is crucial in the road map for a technological exploitation of the novel topological phases of matter. As a matter of fact, it is now well established that TSS exist in crystals and films possessing high structural quality (both stoichiometric/ composition and crystallographic). This raises new challenges in the synthesis of TIs [22]. However, the current progress in growth techniques [23] indicates that TIs are now mature for technology transfer.

The coexistence of several privileged conditions for applications—fast charge carriers [24, 25], sensitivity to applied fields [26], reduced fluctuation of mobility [12], and robustness to disorder [27]— makes TIs promising for technological use in quantum computing [28], high-speed electronic devices [29], as well as spintronic devices with novel functionalities [30]. Moreover, the amazing optical properties of TIs [31-37] allow their versatile and multifunctional use in signal emission, transmission, modulation, and detection.

All these fields of applications benefit from the exploitation of both propagating and localized plasmonic modes in TIs, as achieved for plasmonics with noble metals[38, 39] and graphene [40-44].

Herein, we review applications based on plasmons supported by TIs in the fields of optoelectronics, photonics and thermoplasmonics.

**Terahertz (THz) plasmonics and photodetectors**

The two-dimensional electron gas (2DEG) formed by surface-state electrons support a Dirac plasmon [45-47] with energy $\omega_p$, whose dispersion in the local approximation follows [48]:

$$\omega_p = \sqrt{\frac{D}{2\varepsilon_0 \epsilon} q},$$

(1)

where q is the momentum. The vacuum permittivity and the relative dielectric constant are represented in (1) by $\varepsilon_0$ and $\varepsilon$, respectively. D is the Drude weight [49], which in a 2DEG with particle density n and electron mass $m_e$ is defined as:

$$D = e^2 n/m_e \quad (2)$$

with e the electron charge.

The dispersion relation in Equation (1) yields the square-root dispersion with the momentum q predicted by Stern for a 2DEG [50].

Di Pietro et al. [47] measured the Dirac plasmon in $Bi_2Se_3$ TI with infrared spectroscopy. Plasmons cannot be directly excited by light in smooth samples[51], since photon momentum is always minor than that of plasmons, but the momentum mismatch can be compensated by grating [52]. Therefore, in order to let the light to couple with plasmons, Di Pietro et al. [47] fabricated thin micro-ribbon arrays of $Bi_2Se_3$ with different width to change the plasmon wavevector q up to 0.00015 Å$^{-1}$. An extension of the momentum range up to ~0.3 Å$^{-1}$ has been achieved by exciting the Dirac plasmon with probing electrons [45, 53].

Successively, the plasmonic response of ring structures patterned in $Bi_2Se_3$ films has been studied by THz spectroscopy [54]. The rings exhibit a bonding and an antibonding plasmon modes, whose frequency can be tuned by changing their diameter. The bonding plasmon strongly couples with an optical phonon at about ~2 THz, leading to Fano profiles in the measured extinction spectra.

The most promising application of THz plasmonics with TIs is related to the rectification of THz radiation via the excitation of plasma waves in the active channel of antenna-coupled field-effect transistors (FETs). This photodetection mechanism, proposed by Dyakonov and Shur [55-57], is based on the fact that a FET hosting a 2DEG can act as a cavity for plasma waves, which are launched at the source by means of a modulation of the potential difference between gate and source. Plasma waves propagating in the active channel of FETs cannot be merely recognized with the plasmonic resonance of a 2DEG (2D plasmon) because of the presence of a metal gate [58]. When $q<1/d$ (with d the distance between the 2DEG and the gate), the dispersion relation is modified into $\omega_p = sq$ (at T=0 K and neglecting friction and viscosity) with s the plasma-wave group velocity. Such a linear dispersion relation resembles that of acoustic surface plasmons [59] and sound waves [58].

If a plasma wave reaches the drain contact in a time inferior with respect to the momentum relaxation time, constructive interference in the cavity allows frequency-resolved detection of the incoming radiation ("resonant regime"). In such regime, the dc photoresponse is characterized by peaks at the odd multiples of the lowest plasma-wave frequency. For typical device lengths and carrier densities, the fundamental frequency of plasma waves is in the THz range, so that photodetectors based on the Dyakonov-Shur mechanism are used for THz photodetection.

FET THz detectors conventionally operate at room temperature in the overdamped plasma waves regime, since the length of the FET channel is bigger than the propagation length of plasma waves at room temperature. The electromagnetic ac field, coupled to the source (S) and gate (G) electrodes, simultaneously modulates the carrier density and drift velocity. The resulting current exhibits a dc component, whose magnitude is proportional to the intensity of the incoming radiation and can be measured at the drain (D) contact either in a short circuit (photocurrent mode) or in an open circuit (photovoltage mode) configuration.

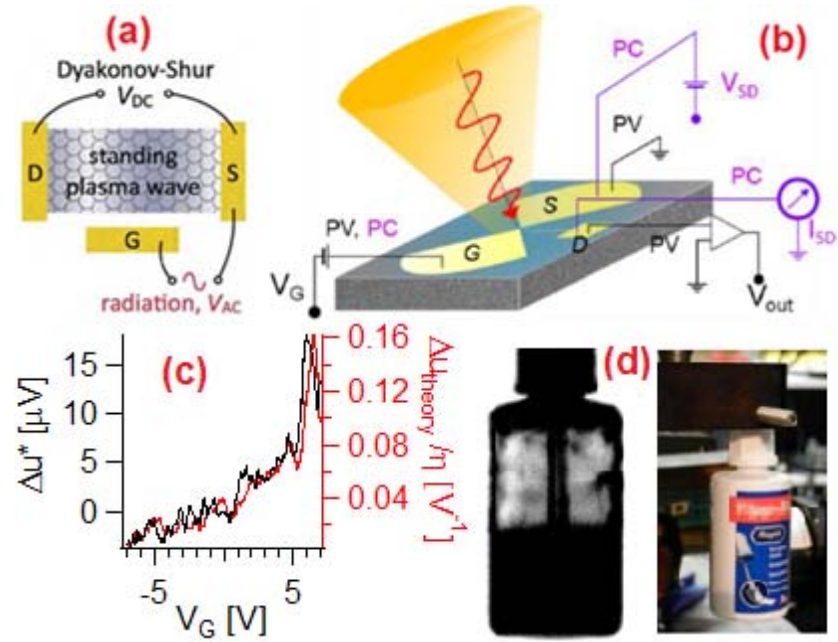

Figure 1: In the plasma-wave-assisted photodetection mechanism, named "Dyakonov-Shur"[56], a finite d.c. voltage is generated in a FET in response to an oscillating radiation field (panel a, adapted with permission from Ref. 60, © 2014 Nature Publishing Group). A sketch of the THz detection principle, photovoltage and photocurrent modes are depicted in panel b in black and purple, respectively (reprinted with permission from Ref. 61, © 2016 American Chemical Society). In panel c, the left panel reports the difference between the source-drain currents measured with the THz beam focused on the sample ($I_{SD,on}$) and without it ($I_{SD,off}$), multiplied by the channel resistance. In the right vertical axis the extrapolated photovoltage ($\Delta u_{theory}$), as predicted by the Diakonov-Shur overdamped plasma-wave theory (Equation 3), divided by the antenna coupling constant ($\eta$). Panel d reports the image recorded by focusing a 0.33 THZ beam on a glue jar. The transmitted power was detected in photovoltage configuration with all electrodes unbiased. The 400 × 700 pixel image, revealing the jar content, was acquired with a time constant of 20 ms. Adapted with permission from Ref. 61, © 2016 American Chemical Society).

Resonant plasma-wave photodetection of THz radiation in graphene FET was demonstrated by both experimentalists [60, 62] and theoreticians [58]. Recently, Vitiello and coworkers[61] demonstrated that TI-based FETs can be even more efficient than graphene, as indicated by the obtained value of the noise-equivalent power (NEP).

According to the Dyakonov-Shur theory[63], in the case of non-resonant (overdamped) detection, the photovoltage Δu follows:

$$\Delta u \propto \eta \frac{1}{\sigma} \frac{d\sigma}{dV_G} \cdot \frac{1}{1 + \frac{R_{ch}}{Z_L}} \qquad (3)$$

where η represents the antenna-dependent coupling efficiency of the incoming radiation [62], σ is the channel conductivity, $R_{ch}$ is the total S-to-D resistance and $Z_L$ is the complex impedance of the readout circuitry. Equation (3), applied to the up-sweeps for a $Bi_2Te_{2.2}Se_{0.8}$-based FET results in the predicted photovoltage trends shown on the right vertical axis of panel c of Figure 1. The comparison with the corresponding experimental data, reported on the left vertical axis, allows asserting that the predominant photodetection mechanism is that related to the excitation of plasma waves.

FET-based THz detectors open new possibilities of construction of real-time THz imaging systems. The result of a transmission imaging experiment of a glue jar only partially filled by using a $Bi_2Te_{2.2}Se_{0.8}$-based FET is shown in panel d of Figure 1. The level of the liquid content of the jar is revealed by THz imaging with a signal-to-noise ratio of ~1000.

**Spin-plasmons**

Due to the spin-momentum locking [64, 65], plasmons in TIs are always accompanied by transverse spin oscillations [66]. The novel spin-plasmon mode has the peculiarity that density fluctuations induce transverse spin fluctuations and vice versa [67]. In a spin-polarized 2DEG, a random-phase approximation model[68] predicts the spin-plasmon lifetime to be sufficiently high in order to enable their technological exploitation in spin-wave-generating devices, such as spin-torque oscillators. The influence of spin-plasmons has been invoked to be crucial in the coupling of Dirac-cone electrons of TIs with phonons[69]. However, attempts to directly probe spin-plasmons by means of electron energy loss spectroscopy have been unsuccessful[53], since the predominating mode in the plasmonic spectrum is a surface plasmon arising from the bulk, free carriers [45, 53]. This mode hybridizes with the spin-plasmon for momenta far from the optical limit[45].

To selectively excite the spin-plasmon, a mechanism exploiting optical spin injection has been proposed by Raghu et al. [67] (Figure 2). These authors suggested that two cross-polarized optical beams of laser pulses can generate a transient spin grating on the surface, whose period is determined by the wavelength and the angle between the two incident beams. Spin-plasmon can be excited whenever the spin grating period matches the plasmon momentum [70, 71]. Experiments reporting (i) the existence of selection rules ruling spin-dependent optical transitions in TIs [72] and (ii) helicity-dependent photocurrents [73] have supported the viability of optical spin injection. However, calculations including both spin-orbit coupling and Zeeman coupling demonstrated [74] that helicity-independent photocurrent dominates over helicity-dependent contributions.

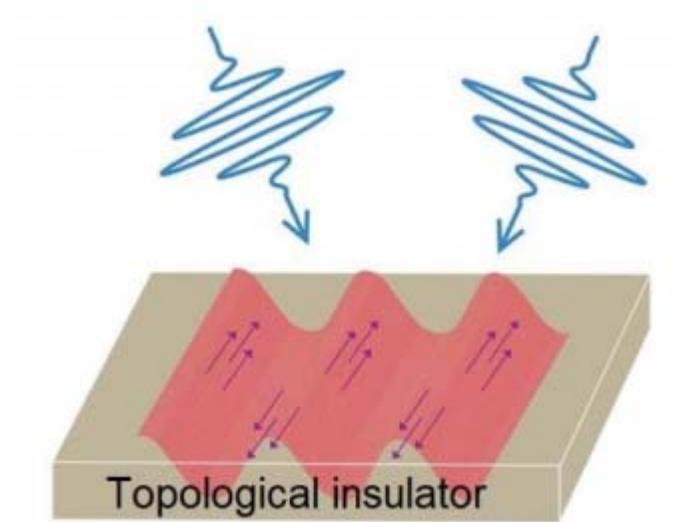

Figure 2: Proposed mechanism to excite spin-plasmons by optical spin injection. The two blue arrows represent impinging optical waves with identical energy. The red wave represents the spin grating, where the arrows specify the direction of the spin. Taken from Ref. 75.

**Magnetoplasmonics**

Recently, magnetoplasmonics is attracting considerable interest for the intriguing prospect of its technological applications[76, 77]. The 2D magnetoplasmons are collective excitations between Landau levels[78] due to electron-electron interactions, which can be observed through infrared optical absorption and inelastic light scattering [79-84].

Autore et al. [85] reported a study on plasmon excitation in an array of Bi$_2$Se$_3$ microribbons for THz light perpendicularly polarized to ribbons at 1.6 K (Figure 3) . Both the collective (plasmon) and the single particle (Drude) excitations have been tuned by an external magnetic field B ranging from 0 to 30 T.

At finite B, plasmon gives rise to a magnetoplasmon mode, whose frequency $\omega_{mp}$ follows:

$$\omega_{mp} = \sqrt{\omega_p^2 + \omega_c^2} \quad (4)$$

where $\omega_c$ is the cyclotron frequency $\pm eB/m_c c$, defined as positive for electrons and negative for holes, with $m_c$ the cyclotron mass and c the speed of light.

The Drude term becomes a cyclotron resonance at finite energy even for low values of B. For magnetic fields higher than 10 T, the cyclotron resonance and the plasmonic modes merge into a unique Dirac magnetic excitation (Figure 3).

Therefore, in TIs, it is possible to achieve a magnetic control of plasmonic modes, paving the way toward THz magnetoplasmonics.

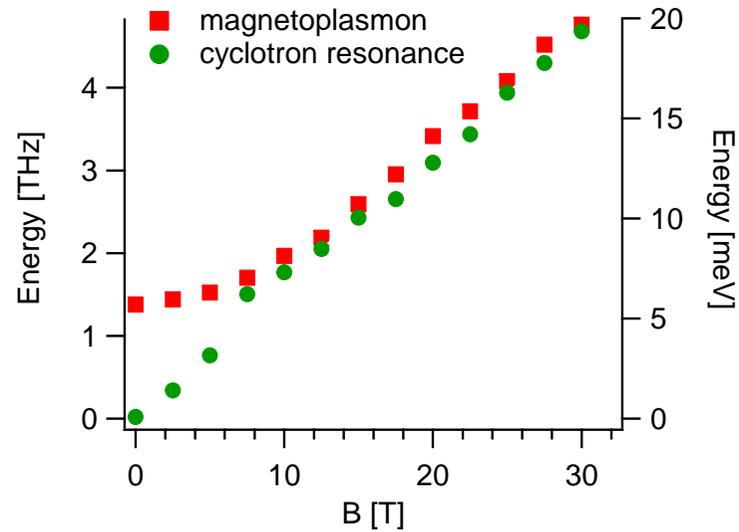

Figure 3. Experimental magnetoplasmon (red squares) and cyclotron (green circles) frequencies as a function of the magnetic field. Adapted from Ref. 85.

**Ultraviolet plasmonics**

Though technological efforts have mainly focused in the spectral range from THz [60] to the visible [41], numerous potential applications of plasmon modes in the ultraviolet (UV) part of the electromagnetic spectrum exist. A possible advantage of UV plasmons is the matching of their high energy with the electronic transition energy of many organic molecules, thus paving the way for UV plasmonics [86], UV imaging [87], DNA sensing [88], UV absorbers [89], and metamaterials with UV plasmonic resonances [90].

Zheludev and coworkers showed that $Bi_{1.5}Sb_{0.5}Te_{1.8}Se_{1.2}$ is a low-loss medium supporting plasmonic excitations in the blue-ultraviolet range [90], in addition to the already-investigated THz frequency range. Metamaterials fabricated from $Bi_{1.5}Sb_{0.5}Te_{1.8}Se_{1.2}$ exhibit plasmonic modes from 350 to 550 nm, while surface gratings show cathodoluminescent peaks from 230 to 1050 nm. The observed plasmonic response (Figure 4) is attributed to the combination of bulk charge carriers from interband transitions and surface charge carriers of the TI.

The plasmonic resonance for perpendicular polarization increases with the length of the groove.

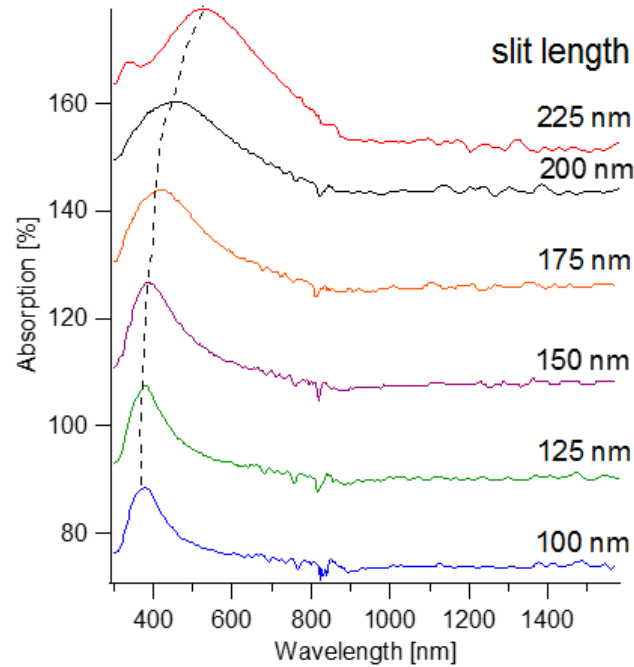

Figure 4: Absorption spectra of various nano-slit arrays of $Bi_{1.5}Sb_{0.5}Te_{1.8}Se_{1.2}$ with lengths D from 100 nm to 225 nm for light polarized perpendicular to the nano-slits. The wavelength of the plasmonic resonance increases with the slit length D. Adapted from Ref. 90.

**Plasmons in topological crystalline insulators (TCIs)**

In a TCI, the space group symmetries of a crystal replace the role of time-reversal symmetry in an ordinary TI (see Refs. 91-95 for more information on TCI). Plasmons of TCI SnTe with nanostructured patterns have been investigated by Wang et al. [96]. Four plasmon resonances excited on the TCI SnTe nanogratings are found in the visible-near-infrared (vis-NIR) spectral region. By variating the grating heights, periodic shifts of resonance wavelengths are observed.

**Thermoplasmonics**

Thermoplasmonics [97-100] is based on the thermal heating originated by the optically resonant excitation of plasmons in nanoparticles [101-103]. Localized surface plasmons (LSP), i.e. plasmon modes in nanostructures, notably enhance the local electric field near the surface of the nanostructures, so as to achieve an enhanced optical response. Photothermal effects activate nanoscale thermal hotspots by light irradiation [104, 105], where plasmon energy is transformed into heat, which increases the temperature of the surrounding medium [106].

Guozhi et al. [107] reported the observation of thermoplasmonic effects due to the excitation of LSPs in $Bi_2Se_3$ TI hierarchical nanoflowers, consisting of a several $Bi_2Se_3$ nanocrystals. In addition, the heat obtained by irradiating $Bi_2Se_3$ nanoplates with a laser in the NIR has been exploited by Li et al. [108] in the fields of cancer imaging and therapy.

**Photonic TIs**

The emerging field of topological photonics[109] promises to introduce into optical physics the various topological phenomena in condensed matter physics, such as the quantum Hall effect. In addition, topological photonics aims to protect photons from undesirable random scattering in their transport from one place to another, similarly to the topologically protected transport of electrons in condensed-matter TIs.

Using appropriately devised electromagnetic media (metamaterials) it is possible to realize topologically non-trivial photonic states, similar to those that have been identified for solid-state TIs. Photonic TIs could have a major impact on optical devices (couplers, waveguides, and so on), making them (i) more robust against scattering from defects or disorder and (ii) more energy efficient.

Recently, photonic topological transport has been reported in many innovative photonic systems [110, 111]. Firstly, quantum-Hall-type phenomena in optics have been predicted by the groups of Raghu & Haldane[112, 113] and Soljačić [114]. The first experimental realization was obtained in the microwave frequency range by means of magneto-optic materials [115]. Nevertheless, magnetic effects are too weak at optical frequencies to achieve photonic TIs with scatter-free edge states. However, it is now demonstrated that it is possible to obtain spin-polarized one-way photon transport without applying external magnetic fields, as shown by Khanikaev et al. [111] by using metacrystals, which consist in superlattices of metamaterials with appropriately designed properties.

Floquet TIs represent a suitable alternative to obtain helical edge states. In Floquet TIs, the temporal modulation of a photonic crystal breaks the time-reversal symmetry, so as to induce topological edge states. Floquet TI are based on an array of evanescently coupled helical waveguides arranged in a honeycomb lattice [110], as sketched in Figure 5. The lattice is needed in order to support Dirac points (a prerequisite for topological phenomena), whereas the helicity of the waveguides disrupts the degeneracy between clockwise and counterclockwise diffraction near the array.

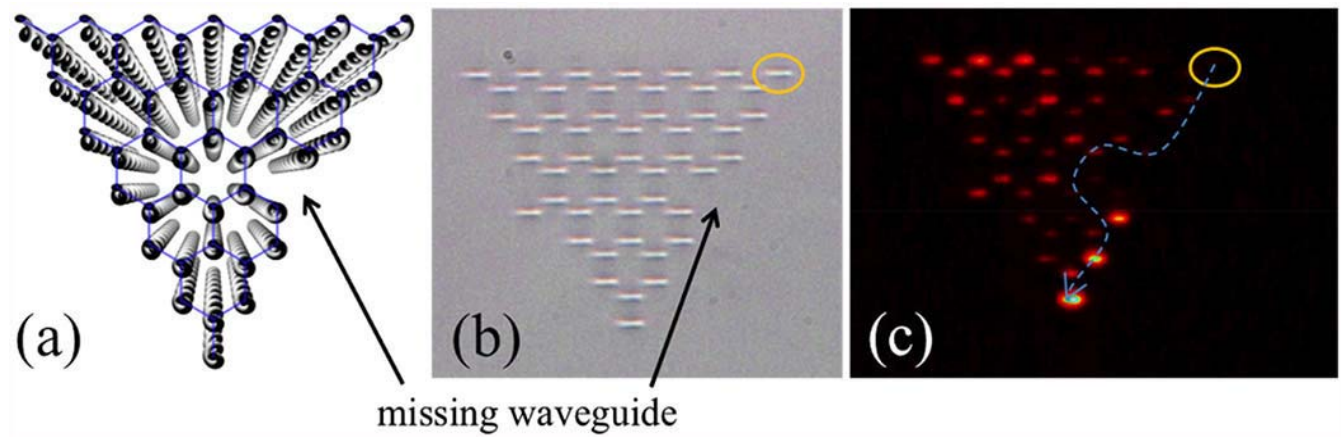

Figure 5: Floquet TIs. (a) Honeycomb lattice of helical waveguides with a 'defect', i.e. an absent waveguide on the edge. (b) Microscope image of waveguide array input facet. (c) Experimental results indicating light input on the top-right waveguide traveling past the defect without backscattering along the edge. Adapted with permission from Ref. 110 (© 2013 Nature Publishing Group)

Another intriguing example of photonic TI has been ideated by Liang and Chong [116] by means of a lattice of coupled ring resonators. By changing the inter-ring coupling strength, it is possible to induce a "topological phase transition" from an ordinary insulator to a TI.

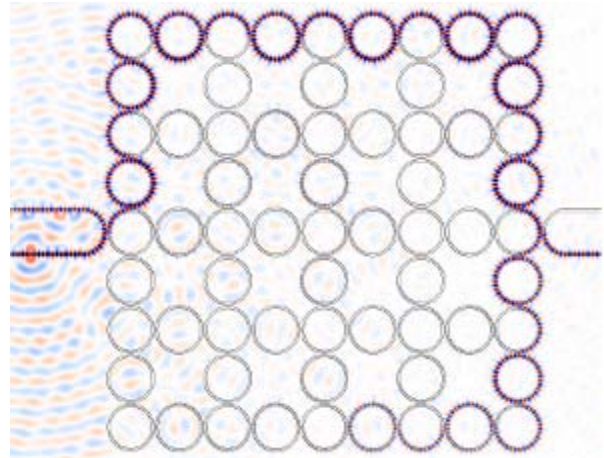

Figure 6: A photonic TI based on a lattice of coupled ring resonators. Light injection occurs on the left (red and blue areas represent the positive/negative values of the electric field). Light is forced to flow around the edge of the lattice, similarly to the case of topological edge states in 2D TIs. Reprinted with permission from Ref. [116] (© 2013 American Physical Society)

**Saturable absorbers**

Bismuth chalcogenides represent promising candidate to generate L-band ultrafast pulses for applications as saturable absorbers [117, 118] for mode-locking. They can be also used as a high-nonlinear medium for mitigating the mode competition of erbium-doped fiber and stabilizing the dual-wavelength oscillation [119]. In many cases, bismuth chalcogenides are used in the form of powders [120] or microfibers [118, 121-125], which do not exhibit

TSS. Therefore, the efficiency of antimony or bismuth chalcogenides as saturable absorbers[118, 121, 124, 126-133] has nothing related to topological phases of matter, in spite of the wealth of claims in literature. Nevertheless, in the following we briefly discuss the most important findings of this research topic.

Li et al. [117] achieved ~360 fs soliton pulse at 1600 nm with a repetition rate of 35.45 MHz by using $Bi_2Se_3$-polyvinyl alcohol film. Using $Bi_2Se_3$ microfibers and with the highest pump power of 5 W, $91^{st}$ harmonic mode locking of soliton bunches with average output power of 308 mW was obtained [118]. In addition, the high third-order nonlinear susceptibility allows the formation of numerous soliton molecules, bound solitons, and soliton rains at a low pump power [122]. At higher values of the pump power, both bunched solitons with soliton number up to 15 and harmonically mode-locked solitons with harmonic order up to 10 were achieved [134].

**Conclusions and Outlook**

The development of disruptive technologies based on the novel topological phases of matter, especially in the field of optoelectronics and photonics, is a great challenge. However, current prototypes of THz photodetectors, photonic devices and thermoplasmonic applications based on TIs have solid possibilities to overcome the up scaling issues.

Even in their first implementation, THz photodetectors using TI-based FETs showed very promising performances. The use of heterostructures of TIs and, moreover, topological Weyl and Dirac semimetals could afford new possibilities for next-generation THz photodetectors. The main target is represented by the design, fabrication and testing of detectors that can be used as pixels in a high-efficiency, high-resolution array for THz imaging.

Furthermore, the physics of photonic topological states is exceptionally promising, with potential implications in various other areas ranging from cold atoms to solid-state electronic systems. New topological phenomena, such as quantum effects with entangled photons and non-equilibrium phenomena, can be explored by optics. Thus, topological photonics is expected to continue engaging researchers in next years, for its intriguing capability to highly improve the performance of optical devices and to limit power consumption.

Future challenges of topological photonics are related to the observation of the nonlinear effects of topological systems, of topological pumping and Bloch oscillations in photonic TIs with novel implementations of one-way waveguiding.

Moreover, the facile control of the plasmons of TCI nanopatterns in the vis-NIR spectral region may have several potential applications. Finally, novel applications could come from thermoplasmonics, which is now ready to concretize the pioneer attempts in cancer therapy.

**Acknowledgements**

This work was partly supported by the European Union ERC Consolidator Grant SPRINT (681379)